\documentstyle[12pt,epsfig]{article}
\begin{document}

\title{Clustering and synchronization with positive Lyapunov exponents}
\author{R. Vilela Mendes \\
Grupo de F\'{\i }sica Matem\'{a}tica,\\
Complexo Interdisciplinar, Universidade de Lisboa,\\
Av. Gama Pinto, 2 - P1699 Lisboa Codex, Portugal}
\date{}
\maketitle

\begin{abstract}
Clustering and correlation effects are frequently observed in chaotic
systems in situations where, because of the positivity of the Lyapunov
exponents, no dimension reduction is to be expected. In this paper, using a
globally coupled network of Bernoulli units, one finds a general mechanism
by which strong correlations and slow structures are obtained at the
synchronization edge. A structure index is defined, which diverges at the
transition points. Some conclusions are drawn concerning the construction of
an ergodic theory of self-organization.
\end{abstract}

\section{Introduction}

Even simple one degree of freedom systems may display a rich dynamical
behavior, namely, sensitive dependence to initial conditions (chaos),
periodic and aperiodic orbits of all types, mixing properties, etc. When
these systems are coupled, their cooperative behavior reveals a set of
dynamical patterns of which the most interesting ones are clustering,
coherent structures and synchronization. The cooperative effects of simple
coupled systems and their dependence on the intensity of the coupling seem
to provide the dynamical basis for many phenomena in physics\cite{Kaneko1} 
\cite{Cross}, chemistry\cite{Kuramoto} \cite{Parmananda} and the
neurosciences\cite{Mirollo} \cite{Skarda} \cite{Singer}. It has also been
suggested by several authors\cite{Robinson} \cite{Hatcher} \cite{Delgado}
that self-synchronized activity in insect colonies has adaptive advantages
and is responsible for efficient task fulfillment.

Synchronization of chaotic systems is a very interesting phenomenon that has
been extensively studied\cite{Pecora1} \cite{Pecora2} and which, in addition
to its role in modeling natural systems, may have technological
applications, for example in the field of secure communications. According
to the theory developed by Pecora and Carrol a subsystem synchronizes with
another (chaotic) subsystem when the corresponding conditional Lyapunov
exponents are negative. The conditional exponents being bona-fide ergodic
invariants\cite{Vilela1}, this is a precise mathematical condition for
synchronization on the support of the invariant measure where the exponents
are defined.

In the standard synchronization scenario, one of the chaotic subsystems
enslaves the other and there is in an effective dimension reduction in
phase-space. However it has been noticed by several authors that there are
situations where one obtains synchronization even with positive conditional
exponents\cite{Shuai} and clustering or strong correlations of the
subsystems even when they are desynchronized\cite{Kaneko2} \cite{Perez} \cite
{Pikovsky}. These correlations have been called {\it hidden coherence}.
Synchronization with positive conditional exponents has been attributed to
the extreme trap effect\cite{Shuai}, namely the fact that near an extreme
point of the iteration function the linear terms vanish and second order
terms may have an effective contracting role. On the other hand fluctuations
in the mean field felt by each individual subsystem and instability of the
solutions for an effective one-dimensional Perron-Frobenius equation, have
been proposed\cite{Kaneko3} as an explanation for the hidden coherence.
These mechanisms may of course play a role in the formation of coherent
structures and specific dynamical features must surely have to be taken into
account for the concrete interpretation of each particular case. However one
would like to understand why clustering and correlation effects are so
common in coupled situations where naively we would still expect to have
very small or no dimension reduction in the overall dynamics.

The method to be used in this paper is also to study a concrete example but
one that is sufficiently simple for almost everything to be exactly computed
and from which essential features may be isolated from model details. In
particular, by using piecewise linear maps one gets rid of second order
effects and non-uniform hyperbolicity. Also the fluctuations in the mean
field, seen by each element of the coupled system, seem to be rather tame
and the origin of the correlations and structures may be correctly
identified. Once the behavior of these phenomena is clearly understood in
this model we will then attempt to separate what seems to be universal
features and what are particular features of the model.

\section{Correlations at the synchronization edge}

Consider a globally coupled system of $N$ Bernoulli units with dynamics 
\begin{equation}
x_{i}(t+1)=(1-c)f(x_{i}(t))+\sum_{j\neq i}^{N}\frac{c}{N-1}f(x_{j}(t))
\label{2.1}
\end{equation}
\ and $f(x)=2x$\ (mod. $1$). The nice feature of this globally coupled
system is that, although each isolated unit is mixing and has orbits of all
types, nevertheless almost everything in the coupled system may be exactly
computed. This avoids interpretation ambiguities of the results and,
hopefully, will allow for the separation of universal features from those
that are model-dependent (see the conclusions). Except for $c=c_{s}=\frac{1}{%
2}\frac{N-1}{N}$ the system is uniformly hyperbolic and the Lyapunov
exponents are: 
\begin{equation}
\begin{array}{lllll}
\lambda _{1} & = & \log 2 &  &  \\ 
\lambda _{i} & = & \log \left( 2\left( 1-\frac{N}{N-1}c\right) \right)  & 
\textnormal{with multiplicity} & N-1
\end{array}
\label{2.2}
\end{equation}

For coupling strength $c<c_{s}$ the Lyapunov dimension is $N$ and one
expects to have a $BRS$-measure absolutely continuous with respect to the
Lebesgue measure in $R^{N}$. This is indeed so, the distribution of the
values taken by any one unit $x_{i}$ is essentially flat and, for large $N$,
the mean field seen by any one unit has very small fluctuations. However as
one approaches $c=c_{s}$ from below, one sees that the dynamics organizes
itself into synchronized patches, with each patch maintaining also an
approximately constant phase relation with the other patches. The
synchronization and phase locking effects however are not absolutely stable
phenomena, the composition and phases of the clusters changing in time but
at a very slow time scale. This clustering effects are evident on the
statistics of the coordinate differences $\left| x_{i}-x_{k}\right| $ shown
in Fig.1, where one has taken the time averages over all pairs of units for
a 100-units system. Fig.1 shows these distributions as $c$ varies. In
Fig.1a, without interaction ($c=0$) the triangular form of the distribution
only reflects the projection along the diagonal of a uniform distribution of
the two coordinates on the unit square. As $c$ increases this distribution
is deformed (Fig.1b, for example) but nothing dramatic occurs until one
reaches the region near $c_{s}=\frac{1}{2}\frac{N-1}{N}$ . Then as shown in
Fig.1c, well defined structures develop which correspond either to
synchronization (peak at zero) or to approximate phase locking (peak near
0.5 and bump around 0.25). Notice however that this figure actually shows
the superposition of two effects. If instead of taking the average over all
distance pairs, one fixes a specific pair of coordinates, one may obtain for
short periods either the two peaks at zero and 0.5 or the bump at 0.25.
Finally for $c>c_{s}$ one obtains global synchronization (Fig.1d). Marked
structures are also obtained for some other linear combinations of the
coordinates. Fig.2a-d shows the statistics for $x_{i}+x_{i+1}-2x_{i+2}$ .

The interpretation of these effects follows nicely from the knowledge of the
Lyapunov exponents listed in (\ref{2.2}). For $c<c_{s}$ all Lyapunov
exponents are positive. However, near $c_{s}$ there is one large Lyapunov
exponent whereas all the others are nearly zero. This implies a fast {\it %
separation dynamics }(sensitive dependence to initial conditions) in one
direction and very slow separation dynamics in all other directions
transversal to the fast one. The fast one corresponds to the eigenvector $%
(1,1,1,1,...,1)$. Therefore, although the invariant physical measure is
still absolutely continuous with respect to Lebesgue, the slow separation
dynamics in the transversal directions corresponds to long wavelength
effects in phase space that are most sensitive to the boundary conditions
and the available phase-space. Then, the slow temporal structures beget
non-uniform probability distributions in the linear combinations of the
variables that correspond to the slow eigenvalues. In particular, $%
x_{i}-x_{i+1}$ corresponds to the eigenvector $(0,...,1,-1,0,...,0)$ and $%
x_{i}+x_{i+1}-x_{i+2}$ to $(0,...,1,1,-2,0,...,0)$.

In conclusion: the existence of structures near the transition points where
one or more Lyapunov exponents approach zero from above should be an
universal phenomena, whereas the detailed form of the structures must depend
one the particular nature of the available phase-space. The non-universality
of the shape of the probability distribution is quite apparent in our
example. $BRS$-measures are in general obtained from the topological
pressure when the function is the sum of the positive Lyapunov exponents,
namely\cite{Gaspard} 
\begin{equation}
\mu _{\phi }(dX)=\lim_{\varepsilon \rightarrow 0}\lim_{T\rightarrow \infty
}\sup_{S}\frac{1}{Z(\varepsilon ,T,\phi )}\sum_{S}\exp \left(
\int_{-T}^{T}\phi (f^{t}Y)dt\right) \times \frac{1}{2T}\int_{-T}^{T}\delta
(X-f^{t}Y)dtdX  \label{2.3}
\end{equation}
$S$ being a $(\varepsilon ,T)-$separated subset. If the function $\phi $ is
the sum of the positive Lyapunov exponents, in our example it only
contributes a phase factor, constant all over phase-space. Hence, the shape
of the invariant measure depends only on the delta function, that is on the
orbit structure which is determined by the boundary conditions and the
available phase-space.

For the coupled Bernoulli units example the shape of the probability
structures may actually be recovered from an approximate probabilistic
equation. From (\ref{2.1}) it follows that for any two units one has the
following relation 
\begin{equation}
x_{i}(t+1)-x_{k}(t+1)=(1-\frac{N}{N-1}c)\left(
f(x_{i}(t))-f(x_{k}(t))\right)   \label{2.4}
\end{equation}
However (\ref{2.4}) does not define a deterministic dynamical law because it
has two branches in the interval $(0,0.5)$, namely 
\begin{eqnarray}
x_{i}-x_{k} &\rightarrow &\left( 
\begin{array}{c}
2(1-\frac{N}{N-1}c)(x_{i}-x_{k}) \\ 
\textnormal{or} \\ 
(1-\frac{N}{N-1}c)\left( 1-2(x_{i}-x_{k})\right) 
\end{array}
\right) \textnormal{if }(x_{i}-x_{k})<0.5  \label{2.5} \\
x_{i}-x_{k} &\rightarrow &
\begin{array}{l}
(1-\frac{N}{N-1}c)\left( 2(x_{i}-x_{k})-1\right) 
\end{array}
\textnormal{if }(x_{i}-x_{k})>0.5  \nonumber
\end{eqnarray}
Defining $r=\left| x_{i}-x_{k}\right| $ and assigning probabilities $p_{1}$, 
$p_{2}$ and $p_{3}$ to these three branches one may write a probabilistic
version of the Perron-Frobenius equation 
\begin{equation}
\frac{\rho (r)}{1-r}=\sum_{y\in f^{-1}(r)}p_{i}\frac{1}{f^{\textnormal{ }^{\prime
}}(y_{i})}\frac{\rho (y_{i})}{1-y_{i}}  \label{2.6}
\end{equation}
where the sum runs over the three possible inverses of $r$ and the factors $%
1-r$ and $1-y_{i}$ account for the projection along the diagonal on the unit
square. Iteration of this equation with $p_{1}=p_{2}=0.5$ and $p_{3}=1$
shows that in the neighborhood of $\frac{N}{N-1}c=0.5$ one obtains the
observed structures, namely either two peaks at zero and 0.5 or a bump
around 0.25.

\section{Self-organization and the structure index}

In general one calls coherent structure (in a collective system) an
identifiable phenomenon that has a scale very different from the scale of
the components of the system. A structure in space will correspond to a
feature at a length scale larger than the characteristic size of the
components and a structure in time is a phenomenon with a time scale larger
than the cycle time of the individual components. This suggests the
definition of a (temporal){\it \ structure index}

\begin{equation}
S=\frac{1}{N}\sum_{i=1}^{N_{s}}\frac{T_{i}-T}{T}  \label{3.1}
\end{equation}
where $N$ is the total number of components (degrees of freedom) of the
coupled system, $N_{s}$ is the number of structures, $T_{i}$ is the
characteristic time of structure $i$ and $T$ is the cycle time of the
components (or, alternatively the characteristic time of the fastest
structure). A similar definition would apply for a spatial structure index,
by replacing characteristic times by characteristic lengths. In our coupled
Bernoulli units example the characteristic times of the separation dynamics
are the inverse of the Lyapunov exponents and one obtains 
\begin{equation}
\begin{array}{lllll}
S & = & \frac{N-1}{N}\left( \frac{\log 2}{\log 2\left( 1-\frac{N}{N-1}%
c\right) }-1\right)  & \textnormal{for} & \frac{N}{N-1}c<0.5 \\ 
S & = & 0 & \textnormal{for} & \frac{N}{N-1}c>0.5
\end{array}
\label{3.2}
\end{equation}
For $\frac{N}{N-1}c>0.5$ the structure index vanishes because the
synchronized motion is effectively one-dimensional and the characteristic
time of the synchronized motion coincides with the characteristic time of
the individual units. The structure index is zero both for the uncoupled
case and the fully synchronized one and diverges at the synchronization
transition.

In a previous paper\cite{Vilela1}, the self-organization that occurs when
identical dynamical systems are coupled was characterized by ergodic
invariants constructed from the conditional exponents. Namely, a {\it %
measure of dynamical self-organization} was defined by\ 
\begin{equation}
I=\sum_{k=1}^{N}\left\{ h_{k}+h_{m-k}-h\right\}   \label{3.3}
\end{equation}
\ where $h_{k}$ and $h_{N-k}$, the {\it conditional exponent entropies}
associated to the splitting\ $R^{k}\times R^{N-k}$ , are the sums of the
positive conditional exponents 
\begin{equation}
\begin{array}{lll}
h_{k} & = & \sum_{\xi _{i}^{(k)}>0}\xi _{i}^{(k)} \\ 
h_{N-k} & = & \sum_{\xi _{i}^{(N-k)}>0}\xi _{i}^{(N-k)}
\end{array}
\label{3.4}
\end{equation}
the conditional exponents being the eigenvalues of 
\[
\lim_{n\rightarrow \infty }\left( D_{k}f^{n*}(x)D_{k}f^{n}(x)\right) ^{\frac{%
1}{2n}}
\]
where $D_{k}f^{n}$\ is the $k\times k$\ diagonal block of the full Jacobian.

For the coupled Bernoulli units example, for splittings into $1$ and $N-1$
parts, one obtains in the limit of large $N$\ 
\begin{equation}
\begin{array}{ccccc}
I & = & \frac{c^{2}}{1-c} &  & c\leq \frac{1}{2} \\ 
& = & -c &  & c\geq \frac{1}{2}
\end{array}
\label{3.5}
\end{equation}
This quantity is also peaked at the synchronization point, (although finite)
but its interpretation is different from the structure index. The Lyapunov
exponents or the conditional exponents measure the change in the dynamics
that occurs when one makes a small change in the initial conditions.
Therefore a system with large exponents has a large freedom to change its
future state with a small effort at the present time. From this point of
view, $h_{k}$ measures the apparent (from the point of view of unit $k$)
dynamical freedom (or rate of information production) of unit $k$. $h_{N-k}$
has the same interpretation for the system composed of the remaining $N-1$
units. However it is $h$ that defines the actual rate of information
production (or dynamical freedom) of the whole system. Therefore $I$ is a
measure of the {\it apparent} excess of dynamical freedom.

\section{Conclusions}

1. When in an interacting multi-unit system the parameters changes and one
reaches a region where one or more of the positive Lyapunov exponents
approaches zero, the slow separation dynamics along the direction of the
corresponding eigenvectors leads to the development of temporal structures,
but without dimension reduction in phase-space. The structures are expected
to be metastable, but with a time scale much larger than the cycle time of
the individual units. These regions, where the system displays what is
perhaps its most interesting behavior, are located near the transition
disorder-order but still on the disorder side. Emergence of structures at
the Lyapunov exponents transition regions is expected to be an universal
phenomenon, but the detailed nature of the structures must be
model-dependent.

2. When there is a natural limitation on the range of values that the state
variable of each individual unit can take, the coupling must be a convex
coupling like in the Bernoulli units example. Then the convex coupling leads
to an overall contracting effect and Lyapunov transitions are to be expected
when the coupling increases. In spatially extended systems, for example, the
basic interaction law might not change but a change in density implies an
effective coupling increase. Therefore in a evolving system where the number
of individuals changes in time (but the available space remains fixed),
effects of the type described here might be expected to arise when the
population density changes.

That at transition regions between chaos and order, evolving systems display
interesting structured properties had been suggested before by several
authors\cite{Langton} \cite{Kauffman}. Why some natural systems might have
evolved to such narrow regions in parameter space is, to a large extent, an
open question. The density-dependent increase of the effective interaction
and the contracting effect implied by the convex coupling, when the amount
of available phase-space remains constant, is a dynamical mechanism that
might explain, in some cases, the evolution towards the transition regions.

3. Given an invariant measure for the interacting system, the structure
index and the measure of self-organization are both well-defined ergodic
invariants which characterize different aspects of the collective behavior.
They provide a first step towards a rigorous {\it ergodic theory of
self-organization}. In this connection it should be mentioned that a
dynamical measure is not completely characterized by the Lyapunov and the
conditional exponents. Ruelle, for example, has pointed out that the
exponents being obtained as limits of averages, the moments of the
fluctuations around the average are new, independent, ergodic invariants
(unless the fluctuations are Gaussian). Moments are not always a reliable
way to characterize stochastic processes\cite{Lukacs} but large families of
ergodic invariants may be obtained in several other ways, for example from a
variational formulation of the dynamics\cite{Vilela2}.

FIGURE CAPTIONS

Fig.1 - Distribution of the variable $\left| x_{i}-x_{j}\right| 
$ for several values of the coupling parameter $C=\frac{N}{N-1}c$ in 
a Bernoulli network of $100$ units

Fig.2 - Distribution of the variable $x_{i}+x_{i+1}-2x_{i+2}$ for 
several values of the coupling parameter $C=\frac{N}{N-1}c$ in a 
Bernoulli network of $100$ units

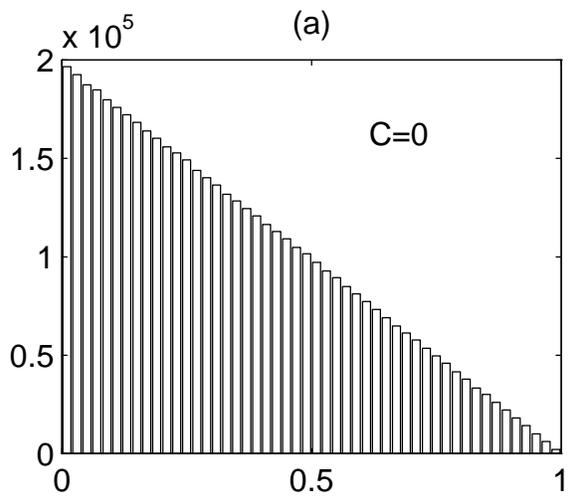
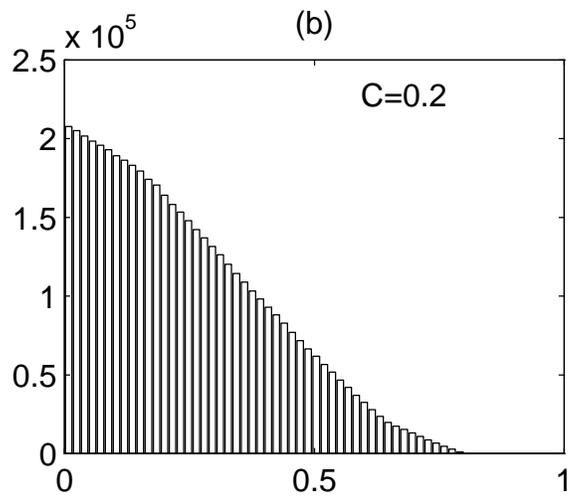
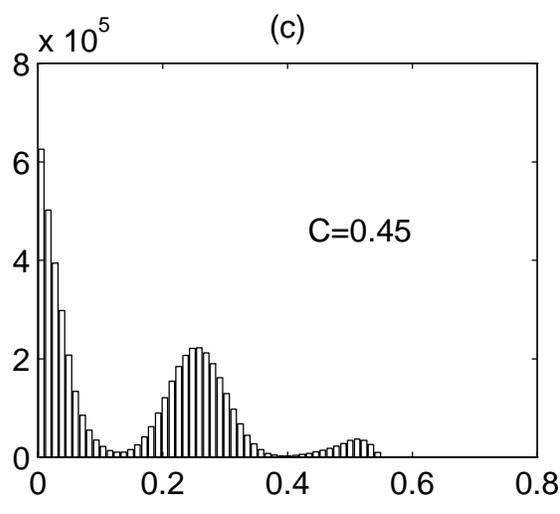
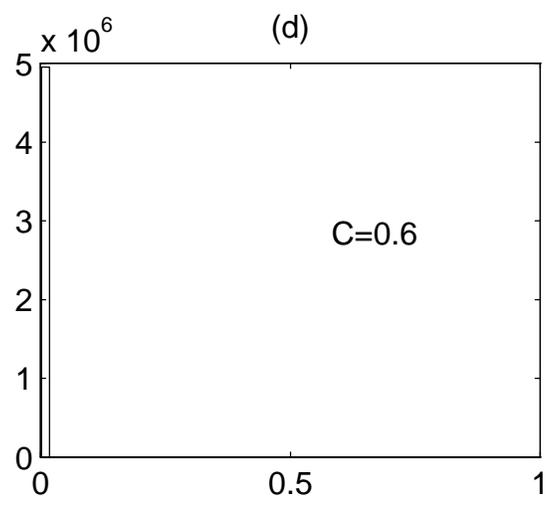
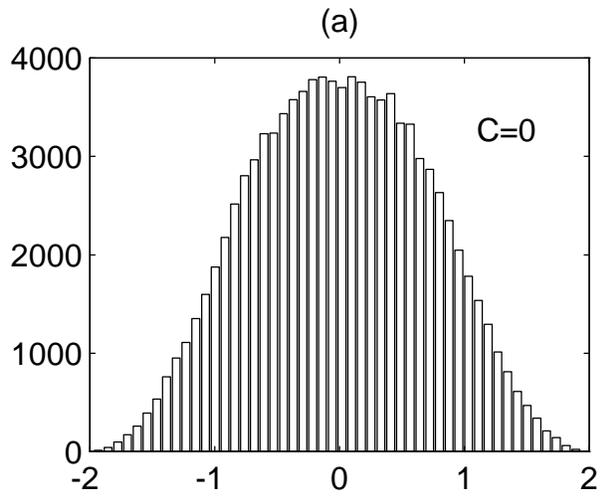
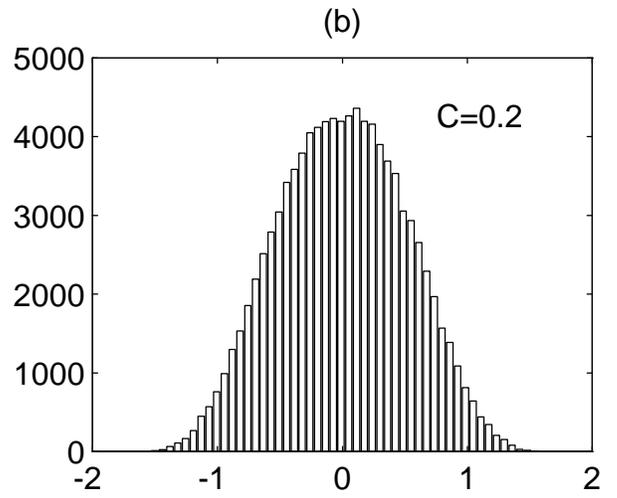
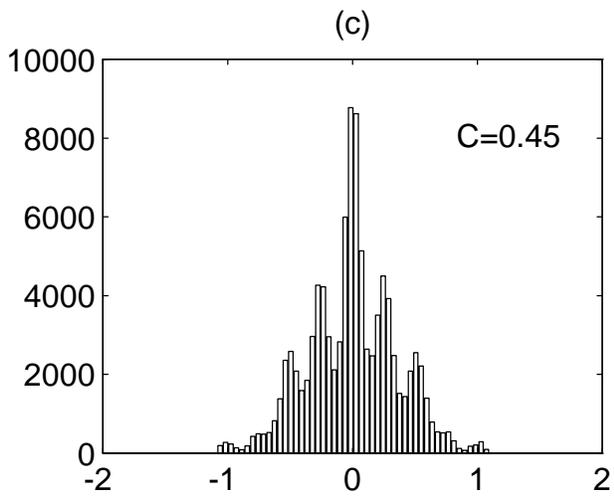
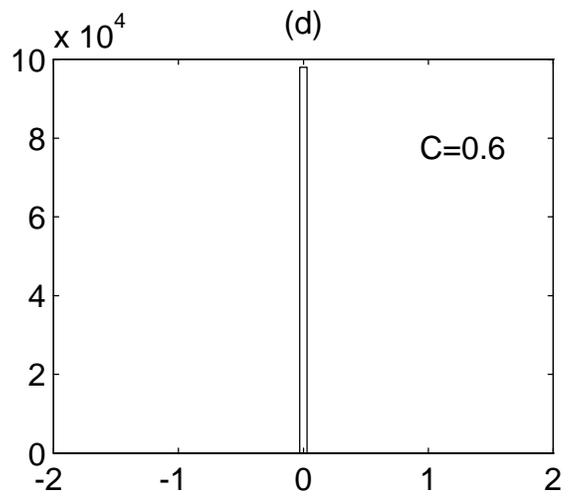
\end{document}